\documentstyle[sprocl]{article}
\def\Journal#1#2#3#4{{#1} {\bf #2}, #3 (#4)}
\def\NPB{{\em Nucl. Phys.} B}
\def\PLB{{\em Phys. Lett.}  B}

\def\ZPC{{\em Z. Phys.} C}
\def\PR{\em Phys. Rev.}

\def\be{\begin{equation}}
\def\ee{\end{equation}}
\def\bea{\begin{eqnarray}}
\def\eea{\end{eqnarray}}

\begin{document}
\title{QCD Laplace Sum Rules and the $\Pi (1300)$ Resonance}
\author{ T.G. STEELE, J.C. BRECKENRIDGE, M. BENMERROUCHE}
\address{Department of Physics and Engineering Physics and,\\
Saskatchewan Accelerator Laboratory\\
University of Saskatchewan\\
Saskatoon, Saskatchewan S7N 5C6, Canada.}
\author{V. ELIAS, A.H. FARIBORZ}
\address{Department of Applied Mathematics, 
The University of Western Ontario, \\
 London, Ontario N6A 5B7, Canada. }
\maketitle\abstracts{
A global fit to the shape of the first QCD Laplace sum rule
exhibiting sensitivity to pion-resonance [$\Pi (1300)$] parameters
is performed, leading to predictions for the pion-resonance mass
and decay constant.  This fit leads to predictions of 
$M_{\pi} = 1.15 \pm 0.15 GeV$ for the pion-resonance mass, and 
$F_\pi^2=(4.5\pm 2.1).(f_\pi^2m_\pi^4/M_\pi^4)$ for the 
pion-resonance decay constant.}

In the work presented here, we use the 
QCD Laplace sum-rules for the longitudinal component of the
axial-vector current correlation function to make explicit predictions
concerning the mass $M_\pi$ and 
the decay constant $F_{\pi}$ of the first excited (pion-resonance) state. 
To leading order in the quark mass m, the QCD Laplace sum rules for
the longitudinal component of the axial vector current correlator
$\Pi^L(s)$ are given by \cite{svz,Dorokhov}
%$^{-}$\cite{Dorokhov}
\begin{eqnarray}
R_0(\tau,s_0) &\equiv& {1\over \pi}\int_0^{s_0} ds \,Im \Pi^L(s)
e^{-s \tau} = -4 m\langle \bar q q\rangle + O (m^2)
\nonumber \\
&\approx&
2 f_\pi^2 m_\pi^2 e^{-m_\pi^2 \tau} + 2 F_\pi^2 M_\pi^2 e^{-M_\pi^2
\tau}
\label{eq:R_0}
\\
       {\cal R}_1(\tau, s_0) 
&\equiv& 
       {1\over \pi}\int_0^{s_0} ds\, Im\Pi^L(s) s \, e^{-s\tau} 
        \nonumber \\
&=& 
m^2 \left( 
           \frac{3}{2\pi^2\tau^2}\left[
                                        1+5\frac{\alpha}{\pi}
                                 \right]
           \left[
                 1-\left(1+s_0\tau\right)e^{-s_0\tau}
           \right]
           -\frac{3}{\pi^2\tau^2}\frac{\alpha}{\pi}\left[
                                                         1-\gamma_{_E}
                                                    \right.
          \right.
\nonumber
\end{eqnarray}
\[
\left. -e^{-s_0\tau}-E_1(s_0\tau)-(1+s_0\tau)e^{-s_0\tau}\log{(s_0\tau)}\right]
-4\,\langle m\bar q q\rangle 
+\frac{1}{2\pi}\langle \alpha G^2\rangle
\]
\[
\left.
+\frac{448\pi}{27}\alpha\left(\langle \bar q q\rangle\right)^2\tau
+{3\rho_c^2\over{2 \pi^2\tau^3}} e^{-\rho_c^2 \over {2 \tau}}
\left[ 
       K_0\left( {\rho_c^2\over {2 \tau}} \right) +
       K_1\left( {\rho_c^2\over {2 \tau}} \right)
\right]
\right)
+O\left( m^3 \right)
\]
\begin{equation}
\hskip -3.5cm \approx
        2f_\pi^2 m_\pi^4 e^{-m_\pi^2 \tau} +
        2F_\pi^2 M_\pi^4 e^{-M_\pi^2 \tau}
\label{eq:R_1}
\end{equation}
\hskip 0.8cm In (\ref{eq:R_0}) and (\ref{eq:R_1}), $\tau$ is the Borel-transform
 parameter, and $s_0$ is
the contin-\\uum threshold for which perturbative-QCD contributions 
to the longitudinal \\
component of the axial vector current correlator
coincide with phenomenological hadronic contributions 
($s_0 > M_\pi^2$).
%\cite{gimenez} 
The parameter $\rho_c$ scales the direct instanton
contribution to the ${\cal R}_1$ sum-rule,\cite{Dorokhov}
which can be excised
simply by going to the $\rho_c \rightarrow \infty$ limit.
%Also, $E_1(x)$ is an exponential integral, \cite{as}
%$\gamma_{_E}\approx 0.5772$ is 
%Euler's constant, and $K_i(x)$ are modified Bessel functions.\cite{as}
It should be noted that a sum-rule analysis containing {\em both} the rather large two-loop
perturbative term  and instanton effects has not previously been performed.
The above sum-rules satisfy a renormalization group (RG) equation 
in $\tau$ which implies that
$m$ and $\alpha$ are the running mass and coupling constant at the energy scale 
$\tau$: \cite{nar}  
\begin{equation}
m(\tau)={\hat m}[-\frac{1}{2}\log{(\tau\Lambda^2)}]^{-4/9}
\quad,\quad \alpha(\tau)=-4\pi/[ 9 \log (\tau\Lambda^2)].
\label{eq:Run_m_alpha}
\end{equation}
The quantity $\hat m$ is RG invariant and is thus a fundamental quark mass 
parameter in QCD. 

The qualitative behaviour of the sum-rules ${\cal R}_0(\tau,s_0)$ and ${\cal R}_1(\tau,s_0)$
provides significant information about the first pseudoscalar resonance.  First, if
$\hat m$ is reasonably small, then ${\cal R}_0(\tau,s_0)$ is essentially independent
of $\tau$, since it is dominated by $\langle m\bar q q\rangle$.  This then implies that the phenomenological
side of the sum-rule is dominated by a light pseudogoldstone boson, since
$\exp{\left(-m_\pi^2\tau\right)}\approx 1$ for 
appropriate mass scales [recall $\tau = 1/M^2$ and note that $\Lambda
< M < {\sqrt s_0} $].  
Thus ${\cal R}_0(\tau,s_0)$ mainly contains the information that the quark condensate 
must balance the phenomenological contribution of the pseudogoldstone
pion, resulting in the GMOR \cite{gellmann} relation
$4m\langle \bar q q\rangle =-2f_\pi^2m_\pi^2$ [ $f_\pi=93 MeV$ ].  

Although the pion dominates ${\cal R}_0(\tau,s_0)$, the excited state $M_\pi$ 
in ${\cal R}_1(\tau,s_0)$ is enhanced relative to the pion by an
additional factor of $M_\pi^2/m_\pi^2$ relative to its contribution
to ${\cal R}_0(\tau,s_0) $. 
For the $\Pi(1300)$ resonance
($M_\pi^2/m_\pi^2\approx 100$), we see that even a 1\% contribution from 
the excited state in 
(\ref{eq:R_0}) corresponds to the excited state's domination of
(\ref{eq:R_1}).

To obtain some quantitative understanding of this excited
pion-resonance state, we utilize the
explicit RG dependence in (\ref{eq:Run_m_alpha}) to obtain from
(\ref{eq:R_1}) an expression whose
purely-perturbative, QCD-vacuum-condensate, and direct instanton contributions 
are independent of the RG-invariant mass $\hat m$:
\begin{equation}
{\cal R}_1 (\tau , s_0)/ {\hat m}^2 \approx
{2f_\pi^2m_\pi^4}/ {\hat m^2}+[{2F_\pi^2M_\pi^4}/ {\hat m^2}]
\exp{\left(-M_\pi^2\tau\right)} 
\label{eq:R_1_on_mhat}
\end{equation}

Known values for the QCD-vacuum condensate permit a least-squares fit 
of the $\tau$-dependence of ${\cal R}_1 (\tau,s_0) /{\hat m}^2 $ to
the form $a\left[ 1+r e^{-M_\pi^2 \tau} \right]$, 
with the fitted parameters $a$, $r$, and $M_{\pi}$ in correspondence
with the right-hand side of (\ref{eq:R_1_on_mhat}): 
$ 
a= 2 f_{\pi}^2m_{\pi}^4 /{\hat m}^2 ; 
 r=F_{\pi}^2
M_{\pi}^4/(f_{\pi}^2 m_{\pi}^4)
$.
It is evident that $\hat m$ only enters the overall normalization of
${\cal R}_1 / {\hat m}^2$ ( the parameter a ).  
The {\it shape} of ${\cal R}_1 / {\hat m}^2$,
while sensitive to $M_{\pi}$ and $F_{\pi}$ (through the parameter $r$
), is entirely decoupled from the RG-invariant quark mass $\hat m$. 

To obtain a fit of this shape dependence to these pion-resonance
parameters, we use the standard set of values $
\langle \alpha G^2 \rangle = 0.045 GeV^4, \hskip 0.3cm \Lambda=0.15 GeV,
\rho_c=1/600 MeV, \alpha \left( \langle \bar q q \rangle \right)^2 = 1.8 \times 10^{-4}
GeV^6$
along with the GMOR relation $\langle m \bar q q \rangle = -
f_{\pi}^2 m_{\pi}^2 / 2 $ with physical values for $m_{\pi}$ and
$f_{\pi}$.
%A $ 50\% $ error is assigned to the total power law contribution, and
%a $15\%$ variation is permitted in the values of $ \rho_c$. The
%weighted least squares fit covers the range $0.5 GeV^{-2} < \tau
%< 1.6 GeV^{-2}$.
The weights for the least squares fit are obtained from a 50\% uncertainty for
power-law corrections, 30\% uncertainty for the continuum hypothesis, and
a 30\% uncertainty for instanton contributions.\cite{Dorokhov}
The optimum value of the parameters
$a$, $r$, $M$ and $s_0$ is then obtained via a fit to the $\tau$ dependence
of $R_1(\tau, s_0)/ {\hat m}^2$ which leads to the 
 smallest value of $\chi^2$.  Uncertainties in the fitted parameters 
 are obtained from
 a Monte-Carlo simulation based upon a 15\% variation in $\rho_c$, 
a possible violation up to a factor of 2 of vaccum-saturation 
for the dimension six condensate, and a simulation of the previously described
uncertainties.

The results of the fitting procedure for 
$ 0.4 GeV^{-2} < \tau < 2 GeV^{-2} $  
are as follows: $M_\pi = 1.15
\pm 0.15 GeV, r=4.5\pm 2.1, a=0.07\pm0.02 GeV^4  $ [uncertainties are at
the 90\% confidence level], with minimum $\chi^2$
occuring at $s_0 = 3.25\pm 1.55 GeV^2$.  The corresponding values of 
the pion-resonance decay constant and RG-invariant quark mass are 
$F_\pi=2.7 \pm 1.3 MeV$ and ${\hat m} = 9.7 \pm 1.4 MeV$,
respectively.  The role of direct instanton contributions can be
understood by comparing to a fit in which such contributions are
absent (the $\rho _c \rightarrow \infty$ limit). Corresponding parameter
values in the $\rho_c \rightarrow \infty $ limit are
$M_\pi = 1.31 \pm 0.16 GeV$,
$r=9.2\pm 3.7$, $a=0.045 \pm 0.015 GeV^4 $, with a minimum $\chi^2$
occuring at $s_0=3.5\pm 1.6 GeV$.  
It is interesting to note that this limit increases the minimum value of 
$\chi^2$.
Thus, the effect of direct
instanton contributions is to lower both the pion-resonance mass and
its decay constant.

\end{document}